\documentclass[prl,a4paper,superscriptaddress,twocolumn]{revtex4-1}
\usepackage{color,amsmath,amssymb,amsthm,times,graphics,graphicx,bm,dcolumn}

\newcommand{\be}{\begin{equation}}
\newcommand{\ee}{\end{equation}}
\newcommand{\ben}{\begin{eqnarray}}
\newcommand{\een}{\end{eqnarray}}
\newcommand{\bes}{\begin{subequations}}
\newcommand{\ees}{\end{subequations}}

\begin{document}

\title{A variational description of the quantum phase transition in the sub-Ohmic spin-boson model}

\author{Alex W. Chin}
 \affiliation{Institut f\"{u}r Theoretische Physik, Albert-Einstein-Allee 11, Universit\"{a}t Ulm, D-89069 Ulm, Germany}

 \author{Javier Prior}
  \affiliation{Departamento de F\'{\i}sica Aplicada, Universidad Polit\'ecnica de Cartagena, Cartagena 30202, Spain}

\author{Susana F. Huelga}
 \affiliation{Institut f\"{u}r Theoretische Physik, Albert-Einstein-Allee 11, Universit\"{a}t Ulm, D-89069 Ulm, Germany}
 \author{Martin B. Plenio}
 \affiliation{Institut f\"{u}r Theoretische Physik, Albert-Einstein-Allee 11, Universit\"{a}t Ulm, D-89069 Ulm, Germany}

\begin{abstract}
The sub-ohmic spin-boson model is known to possess a novel quantum phase transition at zero temperature between a localised and delocalised phase. We present here an analytical theory based on a variational ansatz for the ground state, which describes a continuous localization transition with mean-field exponents for $0<s<0.5$. Our results for the critical properties show good quantitiative agreement with previous numerical results, and we present a detailed description of all the spin observables as the system passes through the transition. Analysing the ansatz itself, we give an intuitive microscopic description of the transition in terms of the changing correlations between the system and bath, and show that it is always accompanied by a divergence of the low-frequency boson occupations. The possible relevance of this divergence for some numerical approaches to this problem is discussed and illustrated by looking at the ground state obtained using density matrix renormalisation group methods.  

\end{abstract} 



\maketitle
The physics of quantum systems in contact with environmental degrees of freedom plays a fundamental role in many areas of physics, chemsitry and biology, including systems as diverse as solid state quantum computers\cite{kit,kit2}, quantum impurities\cite{bullarev08}, and photosynthetic biomolecules\cite{nat2, plenio08, chinnat10,caruso09,prior10}.  A key theoretical model for the study of system-environment interactions is the spin-boson model (SBM), which consists of a two-level system (TLS) that is linearly coupled to an `environment' of harmonic oscillators \cite{leggett,weiss}. Although this model has been studied extensively, there are still many open problems in SBM physics, most notably those concerning the quantum phase transition (QPT) between delocalised and localised phases that exists in the SBM when the oscillators are characterised by sub-Ohmic spectral densities. 

The standard quantum-classical mapping predicts that the sub-ohmic SBM should be equivalent to a classical Ising spin chain
with long-range interactions, and predicts a continuous magnetic transition with mean-field critical exponents for $0<s<0.5$. In Ref. \cite{vojta05}, a continuous transition in the sub-Ohmic SBM was observed using the numerical renormalisation group (NRG) technique for all values of
$0 < s < 1$, but the critical properties of the transition were found to be non-mean-field for $0<s<0.5$. It was suggested that this implied a breakdown of the classical to quantum mapping, and some subsequent work in this and other systems has supported this claim \cite{kirchner09,kirchner10}. However, it is now believed that the non-mean-field results found by NRG in $0<s<0.5$ are incorrect, and arise from the truncation of the number of states $N_{b}$ used to describe each oscillator in the Wilson chain \cite{vojta10,hou10}. Recent studies of the sub-Ohmic QPT using quantum monte carlo (QMC) \cite{winter09}, sparse polynomial space approach (SPSA) \cite{alvermann09}, and an extended coherent state technique have indeed found mean-field critical exponents for $0<s<0.5$ \cite{zhang10}. 

In this article we propose a variational ansatz for the ground state of the sub-Ohmic SBM for $0<s<0.5$ which does not require any truncation of the environment. This is an important feature, as we shall show that the number of environmental  bosons \emph{diverges} above the transition. Although our ansatz is essentially just a guess at the form of the ground state, we shall show that our results agree extremely well with existing numerical results and support the mean-field picture of the transition for $0<s<0.5$. Moreover, we are able to give an inuitive picture of the phase transition that might help in improving numerical simulations of the ground state, and illustrate this with a density matrix renormalisation group simulation of the SBM ground state.
 
\section{The variational ansatz}

The spin-boson Hamiltonian can be written ($\hbar=1$) as \cite{winter09,chinchain10,weiss}

\begin{equation}
H=-\frac{1}{2}\Delta\sigma_{x}+\frac{1}{2}\sigma_{z}\sum_{l}g_{l}(a_{l}+a_{l}^{\dagger})+\sum_{l}\omega_{l}a_{l}^{\dagger}a_{l}.
\end{equation}
where $\sigma_{i}$ are the usual Pauli operators which describe the TLS and $a_{l},a_{l}^\dagger$ are the bosonic anihilation and creation operators respectively of bath modes of frequency $\omega_{l}$. The tunneling amplitude of the TLS is $\Delta$, and $g_{l}$ are the couplings between the TLS and the bath modes. It is well established that all effects of the bath on the reduced state of the TLS are completed determined by the spectral function $J(\omega)=\pi\sum_{l}g_{l}^{2}\delta(\omega-\omega_{l})$ \cite{leggett,weiss}. Following Bulla et al. we consider the spectral function $J(\omega)=2\pi\alpha\omega_{c}^{1-s}\omega^{s}\Theta(\omega_{c}-\omega)$ \cite{Bulla05}, where $\omega_{c}$ is the maximum frequency in the bath. Super-Ohmic baths have $s>1$, Ohmic baths $s=1$ and sub-Ohmic baths $s<1$.

Representing the $+1,-1$ eigenstates of $\sigma_{z}$ as $|+\rangle,|-\rangle$ respectively, a variational ground state ansatz $|\Psi\rangle$ is written in the following way,

\begin{equation}
|\Psi\rangle=C_{+}|+\rangle\otimes|\phi_{+}\rangle +  C_{-}|-\rangle\otimes |\phi_{-}\rangle,
\label{psi}
\end{equation}
where $|\phi_{\pm}\rangle$ is given by
\begin{equation}
|\phi_{\pm}\rangle=e^{-\sum_{l}f_{l\pm}(a_{l}-a_{l}^{\dagger})}|0\rangle,
\end{equation}
and $|0\rangle$ is the vacuum of the bath modes. This ansatz describes a dressed TLS, a superposition of the localised states $|\pm\rangle$ which are correlated with bath modes displaced by $f_{l\pm}$. For general displacements, the wave function is not separable into a product state of the TLS and environment, and our ansatz is not simply a mean field theory like the that recently studied by Hou et al.\cite{hou10}. Our ansatz is a generalisation of the variational wavefunction of Silbey and Harris (SH) \cite{silbey}, in which the constants and displacements are fixed to obey $C_{+}=C_{-}$ and $f_{l+}=-f_{l-}$. As we shall show, these constraints are broken in the localised phase, and we shall henceforth refer to the ansatz of Eq. (\ref{psi}) as the Asymmetrically Displaced Oscillator (ADO) state. The normalisation of the wavefunction $\langle \Psi|\Psi\rangle=1$ enforces the relation $C_{+}^{2}+C_{-}^{2}=1$, and we assume that $C_{+},C_{-}$ are real.  The order parameter of the TLS localisation transition will be taken as the magnetisation $M=\langle \Psi|\sigma_{z}|\Psi\rangle$, which can be expressed as $M=C_{+}^2-C_{-}^{2}$. Using the normalisation condition we obtain $C_{\pm}^{2}=\frac{1}{2}(1\pm M)$. With these relations and the standard properties of displaced oscillators, one obtains the following expression for the ground state energy $E(M)=\langle \Psi|H|\Psi\rangle$,
\begin{eqnarray}
E(M)&=&-\frac{1}{2}\tilde{\Delta}\sqrt{1-M^2}+\frac{(1+M)}{2}\sum_{l}(f_{l+}g_{l}+f_{l+}^{2}\omega_{l})\nonumber\\
&-&\frac{(1-M)}{2}\sum_{l}(f_{l-}g_{l}-f_{l-}^{2}\omega_{l}).
\label{energy}
\end{eqnarray}

In Eq. (\ref{energy}) we have introduced a renormalized tunneling amplitude $\tilde{\Delta}$ given by 
\begin{equation}
\tilde{\Delta}=\Delta\langle \phi_{+}|\sigma_{x}|\phi_{-}\rangle=\Delta\exp\left[-\frac{1}{2}\sum_{l}(f_{l+}-f_{l_{-}})^{2}\right].
\label{tildedelta}
\end{equation}
The quantity $\tilde{\Delta}/\Delta$ is the overlap of the displaced oscillator wave functions which dress the TLS states $|\pm\rangle$. We now minimize the energy at constant $M$ w.r.t. the displacements $f_{i,l}$ to obtain,
\begin{eqnarray}
f_{l\pm}&=&-\frac{g_{l}(M\tilde{\Delta}\pm\sqrt{1-M^2}\omega_{l})}{2\omega_{l}(\tilde{\Delta}+\sqrt{1-M^2}\omega_{l})}.
\end{eqnarray}
These displacements are then substituted back into Eq.(\ref{energy}), and the sums converted into integrals using the spectral function. The integrals can be computed exactly in terms of hypergeometric functions, but in the scaling limit $\omega_{c}\rightarrow\infty$, these hypergeometric functions can be expanded in powers of $\omega_{c}^{-1}$. For $s<0.5$, the ground state energy to leading order in $\frac{\Delta}{\omega_{c}}$ takes the simple form, 

\begin{eqnarray}
E&=&-\tilde{\Delta}\sqrt{1-M^2}-\frac{\alpha\omega_{c}}{2s}\nonumber\\
&+&\frac{\alpha\pi\omega_{c}(1-s)(1-M^2)}{2\sin(\pi s)}\left(\frac{\tilde{\Delta}}{\omega_{c}\sqrt{1-M^2}}\right)^{s}\label{energyfinal}.
\end{eqnarray}
\begin{figure}
\includegraphics[width=7cm]{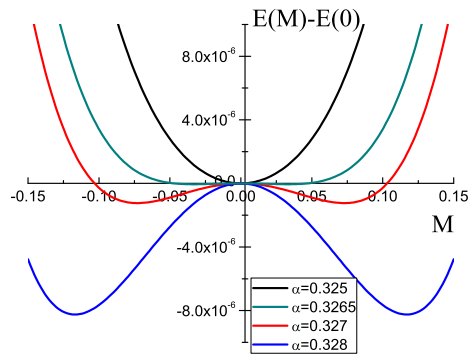}
\caption{ Ground state energy difference $E(M)-E(0)$ as a function of magnetization for $s=0.3$, $\Delta=1$ and $\omega_{c}=10$. The curves correspond to $\alpha=0.325,0.3265,0.327$ and $0.328$ (black,green, red and blue lines respectively).}
\label{energypic}
\end{figure}

\begin{figure}
\includegraphics[width=7cm]{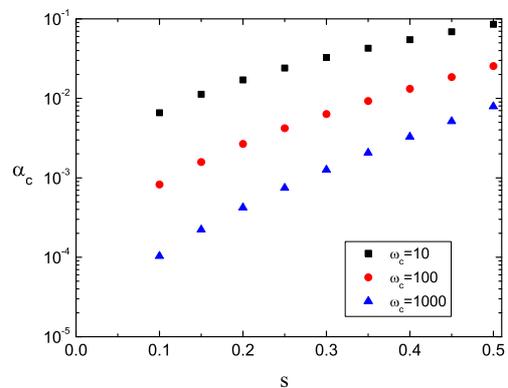}
\caption{Analytical critical coupling $\alpha_{c}$ as a function of $s$ for $\Delta=1$ and various $\omega_{c}$.}
\label{phase}
\end{figure}
The renormalised tunneling amplitude is the solution of the implicit equation,
\begin{equation}
\tilde{\Delta}=\Delta\exp\left[-\alpha\omega_{c}^{1-s}\int_{0}^{\omega_c}\frac{(1-M^2)\omega^{s}d\omega}{(\tilde{\Delta}+\sqrt{1-M^2}\omega)^{2}}\right],\label{ird}
\end{equation}
which in the scaling limit takes the form 
\begin{equation}
\tilde{\Delta}=\Delta\exp\left[\frac{\alpha}{1-s}-\frac{\alpha\pi s}{\sin(\pi s)}\left(\frac{\omega_{c}\sqrt{1-M^2}}{\tilde{\Delta}}\right)^{1-s}\right].
\label{renorm}
\end{equation}
Equation (\ref{renorm}) has a number of self-consistent solutions. For sub-ohmic baths one of these solutions is always $\tilde{\Delta}=0$. This solution corresponds to the complete localisation of the TLS ($M=\pm1$) and is made self-consistent by the infra-red divergence of the integrand in Eq. (\ref{ird}) for $s<1$ and $\tilde{\Delta}=0$. This divergence arises from the divergence of the boson number of the low frequency modes when subject to a static force, which causes $|\phi_{\pm}\rangle$ to become orthogonal \cite{chin06,lu07,leggett,weiss}. As discussed in Refs. \cite{chin06,lu07}, for sufficiently small $\alpha$ there are also finite solutions for $\tilde{\Delta}$, and the physcially relevant one can be can be expressed analytically in terms of the Lambert $W$ function \cite{lambert}. With this solution the groundstate energy can be written as a function of just the original system parameters and the magnetisation. To obtain the ground state magnetisation we then simply minimise the ground state energy of Eq. (\ref{energyfinal}) w.r.t. $M$ for fixed $\alpha,\Delta$ and $\omega_{c}$. 

\section{Ground state energy, critical exponents and critical couplings}\label{results}

The ground state energy can be Taylor-expanded about $M=0$. When this is done we find that for small $M$ the energy takes the Ginzburg-Landau form $E=c_{0}(\alpha)+c_{1}(\alpha)M^2+c_{2}(\alpha)M^4+O(M^6)$ where $c_{i}(\alpha)$ are constants for fixed $\alpha,\omega_{c},\Delta$. This form for the ground state energy guarantees a second-order magnetic transition; above a critical coupling $\alpha_{c}$, a magnetisation grows continuously with $M\propto |\alpha-\alpha_{c}|^{\frac{1}{2}}$ and the magnetic susceptibility $\chi\propto|\alpha-\alpha_{c}|^{-1}$. The typical development of the continuous QPT is shown in Fig. \ref{energypic}. The critical coupling $\alpha_{c}$ is the coupling for which $c_{1}(\alpha_{c})=0$. In the scaling limit this equation can be solved analytically and the critical coupling is obtained as,

\begin{equation}
\alpha_{c}=\frac{\sin(\pi s)}{2\pi(1-s)}\left(\frac{\Tilde{\Delta}_{c}}{\omega_{c}}\right)^{1-s}.
\end{equation}
In order to obtain $\alpha_{c}$ as a function of the original system parameters, the renormalised tunneling matrix element at the critical point $\tilde{\Delta}_{c}$ must also be determined self-consistently from Eq. (\ref{renorm}). Fortunately, in the scaling limit $\tilde{\Delta}_{c}$ can also be found analytically, leading to the final prediction,
\begin{equation}
\alpha_{c}=\frac{\sin(\pi s)e^{-s/2}}{2\pi(1-s)}\left(\frac{\Delta}{\omega_{c}}\right)^{1-s}.\label{critalpha}
\end{equation}
Figure \ref{phase} shows the dependence of the critical couplings obtained from the variational ground state as a function of $s$ for various values of $\omega_{c}$ and $\Delta=1$. Note that the predicted values agree well with those obtained by recent numerical QMC, NRG and SPSA studies, and reproduces the scaling $\alpha_{c}\propto\left(\frac{\Delta}{\omega_{c}}\right)^{1-s}$ previous seen in NRG and other approaches \cite{Bulla05,vojta05,chin06,lu07,alvermann09,winter09}.

\section{Observables}
\subsection{Magnetization and coherence}\label{magnetisation}
\begin{figure}
\includegraphics[width=7cm]{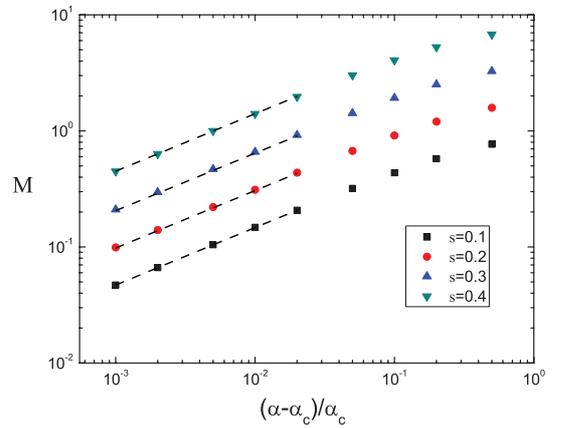}
\caption{Magnetisation $M$ as a function of $(\alpha-\alpha_{c})/\alpha_{c}$ for $\alpha>\alpha_{c}$, $\Delta=1$ and $\omega_{c}=10$. For visibility, the curves have been multiplied by $1,2,4,8$ for $s=0.1,0.2,0.3,0.4$ respectively. Dashed lines correspond to $M\propto \sqrt{\frac{\alpha-\alpha_{c}}{\alpha_{c}}}$.}
\label{mag}
\end{figure}
\begin{figure}
\includegraphics[width=7cm]{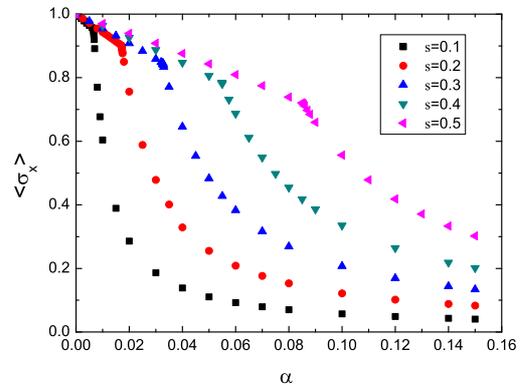}
\caption{ Expectation value $\langle \sigma_{x}\rangle$ as a function of $\alpha$ for $\Delta=1$ and $\omega_{c}=10$.}
\label{sx}
\end{figure}

Using the values of $\alpha_{c}$ given in Eq. (\ref{critalpha}), we show in Fig. (\ref{mag}) that the magnetisation obtained from the energy minimization does indeed behave like $M\propto(\alpha-\alpha_{c})^{\frac{1}{2}}$ close to the transition. The magnetisation data are again in good agreement with the QMC and SPSA results \cite{winter09,alvermann09}. Figure \ref{sx} shows the behaviour of the cohrence $\langle \sigma_{x}\rangle$ as a function of $\alpha$ for $\Delta/\omega_{c}=0.1$. We find that $\sigma_{x}$ is always continuous at the transition, but $\frac{\partial \langle \sigma_{x}\rangle}{\partial \alpha}|_{\alpha_{c}} $ is discontinuous. In the scaling limit the variational theory predicts $\langle \sigma_{x}\rangle=e^{-\frac{s}{2(1-s)}}$ at the critical point, and is thus independent of $\Delta/\omega_{c}$. Above the transition $\langle\sigma_{x}\rangle$ decays faster, but persists well into the localised phase. The persistence and $s$-dependence of $\langle\sigma_{x}\rangle$ at and above the transition point seems to be generally consistent with the dynamical NRG study of Anders et al. \cite{anders07}, where it was found that a coherent tunneling peak in the equilibrium correlation function survives well into the localised phase for $0<s<0.5$ and becomes \emph{more} robust as $s\rightarrow0$. This behaviour was also observed by Anders et al. in the non-equilibirum dynamics, where oscillation amplitudes at criticality become stronger as $s$ descreases. This trend is described in our approach by the \emph{increasing} value of $\langle\sigma_{x}\rangle$ around the transition as $s$ decreases. 

\begin{figure}
\includegraphics[width=7cm]{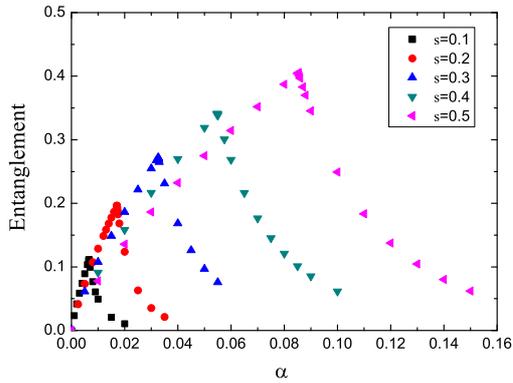}
\caption{Entanglement between the spin and bosonic environment as a function of $\alpha$ for $\Delta=1$ amd $\omega_{c}=10.$}
\label{ent}
\end{figure}
\subsection{Entanglement}\label{entanglement}
In Fig.\ref{ent} the entanglement between the spin and the bosonic environment is plotted as a function of $\alpha$. For globally pure states the entanglement $\mathcal{E}$ is quantified by the entropy of the reduced density matrix of the TLS, and is thus given by $\mathcal{E}=-p_{+}\log(p_{+})-p_{-}\log(p_{-})$, where $p_{\pm}=\frac{1}{2}(1+\sqrt{\langle\sigma_{x}\rangle^{2}+\langle\sigma_{z}\rangle^{2}})$. We observe a very similar cusp-like maximum in the entanglement $\mathcal{E}_{Max}$ at $\alpha_{c}$ to that found in a recent NRG study \cite{lehur07}.  As $\mathcal{E}_{Max}$ is determined only by the value of $\langle\sigma_{x}\rangle$ at $\alpha_{c}$, we also predict that $\mathcal{E}_{M}$ goes to a constant in the scaling limit with $p_{\pm}=0.5(1\pm \tilde{\Delta}_{c}/\Delta)$. The entanglement quantifies the degree of non-classical correlations between the TLS and bath, and we shall now discuss the intutive physcial picture of the transition which arises from a microscopic analysis of the correlations quantitfied by the entanglement.

\section{Discussion}\label{discussion}
\subsection{Delocalised phase}
In the delocalised phase ($M=0$) we find that $f_{l+}=-f_{l-}$ for all modes, and the AOD state thus coincides with the SH variational ansatz. The sub-Ohmic SH state was previously studied in Refs. \cite{chin06,lu07}, and was shown to possess a transition where $\tilde{\Delta}$ jumped discontinuously from a finite value to zero at a critical coupling $\alpha_{c}$. A similar discontinuous transition was also found by flow-equation analysis \cite{kehrein}. However the discontinuous transition is in fact an artefact of the inflexibility of the SH ansatz, as it can only describe a delocalised phase or a completely localised ($\tilde{\Delta}=0, M=1$) state; the discontinuous transition arises simply from the abrupt change from a delocalised to a completely localized state when the latter becomes energetically favourable. 

In the delocalised phase, the physics of the AOD/SH state is determined only by the renormalisation of $\tilde{\Delta}$ by the bath. The variational solution separates the bath into adiabatic modes (A-modes) and non-adiabatic modes (NA-modes) which have very diferrent frequency responses to the renormalised TLS tunneling. For the fast, high frequency A-modes $(\omega_{l}\gg\tilde{\Delta})$, the TLS tunneling appears to be a very slowly varying force, and the A-modes can adibatically adjust their displacements to maximise their interaction energy with the TLS ($f_{l\pm}\approx\pm g_{l}\omega_{l}^{-1}$) \cite{leggett,weiss,chin06}. The slow, NA-modes with $\omega_{l}\ll\tilde{\Delta}$ cannot respond fast enough to follow the tunneling and their displacement is supressed at low frequency ($f_{l\pm}\approx\pm g_{l}\omega_{l}\tilde{\Delta}^{-1}$) \cite{chin06,lu07}.  From Eq (\ref{ird}), one can see that this supressed displacement of NA-modes self-consistently permits a finite solution for $\tilde{\Delta}$ by preventing the infra-red divegence of the integrand in Eq. (\ref{ird}) \cite{chin06,lu07}.  Despite the different frequency response of A and NA-modes, \emph{all} modes are correlated in the same qualitative way with the TLS states ($f_{l+}=-f_{l-}$). Looking at Eq. (\ref{psi}), one sees that with these dressing correlations, the ground state is not separable into TLS and bath variables, and thus there is entanglement between the TLS and bath at finite $\alpha$. As dressing correlations monotonically suppress $\tilde{\Delta}$ as $\alpha$ is increased,  the entanglement also increases monotonically as the transition is approached from the delocalised phase.

\subsection{Localised phase}
Above the transition ($M\neq0$), we find that a new energy scale appears in the problem, and the displacements $f_{l,\pm}$ are no longer of equal magnitude and opposite sign. Modes with $\omega_{l}\gg M\tilde{\Delta}(1-M^2)^{-\frac{1}{2}}$ continue to be adiabatic, whilst non-adiabatic modes with $\omega_{l}\ll M\tilde{\Delta}(1-M^2)^{-\frac{1}{2}}$ now have displacements which have the \emph{same} sign and \emph{grow} at low frequency, $f_{l+}=f_{l-}=-Mg_{l}\omega_{l}^{-1}$. 

Because the non-adiabatic mode displacements have the same sign, they are not correlated with the state of the TLS. As a result, the state of the system is essentially separable, and takes the form of a product state of NA-modes and the correlated TLS-A-mode wave function. The NA-mode wave function has a finite displacement, and this appears to the TLS as an effective `mean-field'-like magnetic field in the $z$-direction. Indeed, the displacement of the NA modes induced by the magnetization of the spin is precisely what one would obtain from a mean field treatment of the problem \cite{hou10}. While the total average displacement of NA modes is finite, there is an infra-red divergence of the occupation number of these modes, as $\sum_{l}\langle a_{l}^{\dagger}a_{l}\rangle_{NA} \propto M^{2}\int_{0}^{\tilde{\Delta}} d\omega \omega^{s-2}$. However, from Eq. (\ref{tildedelta}), we see that the fact that $f_{l\pm}$ have the same sign leads to no supression of $\tilde{\Delta}$ from NA modes. Thus in spite of the infra-red divergence of the low frequency boson number, the coherence of the gound state remains \emph{finite} at the transition and allows $M$ to grow continuously above the transition. 

As the magnetisation increases, the energy scale that divides the NA and A modes also increases, and more of the bath modes become non-adiabatic and uncorelated with the TLS. This trend is reflected in the monotonic descrease of entanglement above the transition, and also in the qualitative change in the oscillator-induced suppression of the TLS coherence. In the localised phase, the suppression of $\langle\sigma_{x}\rangle$ becomes increasingly due to the NA-mode bias, which caues the efective magnetic field seen by the TLS to point away from the $x$ axis. In fact, as $\alpha\rightarrow\infty$ the suppression of $\tilde{\Delta}$ due to dressing vanishes, and the suppression of $\langle \sigma_{x}\rangle$ is determined solely by the rotation of the ground state to lie along the effective NA-mode magnetic field. 

\section{Fidelity of AOD ansatz with DMRG ground state}
\begin{figure}
\includegraphics[width=7cm]{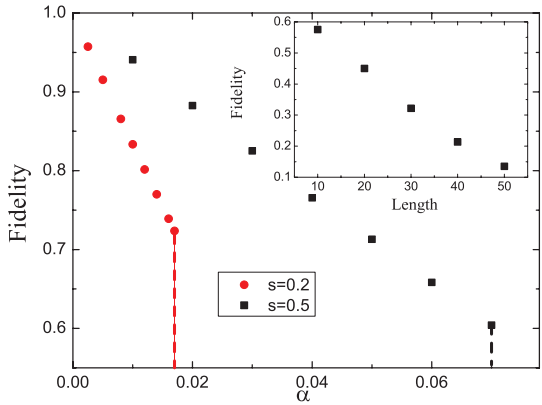}
\caption{Fidelity of the variational ansatz with the ground state determined by DMRG as a function of $\alpha$ for $\Delta=1$ amd $\omega_{c}=10$. The DMRG simulation used $100$ sites, $N_{b}=15$ and $10$ schmidt coefficients were retained. At the critical coupling, the fidelity drops suddenly from finite values to zero. The inset shows the typically dramatic descrease in the fidelity with system size above the transition. Inset data corresponds to $s=0.3$, $\alpha=0.3280$ $ (\alpha_{c}=0.3267)$.}
\label{dmrg}
\end{figure}
While there is general agreement between the observables obtained from the AOD ansatz and previous numerical studies of the transition, it is not posssible in most cases to compute the fidelity of AOD ansatz with the ground states obtained by these methods. In order to explore how close the AOD ansatz is to the true ground state, we simulated the ground state using DMRG methods. When the results converge, the DMRG method produces an extremely accurate matrix product state representation of the ground state wavefunction of the system and bath, and the fidelity (overlap) of the AOD ansatz and the MPS ground state can be easily computed.  To implement the DMRG simulation, the SBM is first mapped onto a semi-infinite chain of harmonic oscillators using an analytical unitary transformation previously used to study open-system dynamics with time-adaptive DMRG (t-DMRG) \cite{prior10}. Unlike the mapping onto a Wilson chain that is used in NRG approaches, our mapping does not require any discretisation of the spectral density of the bath, although our mapping can also be carried out analytically for linear and logarithmically-discretised baths, as shown in Ref. \cite{chinchain10}. Once mapped to a chain, the ground state of the model is obtained using imaginary-time t-DMRG evolution. Figure (\ref{dmrg}). shows some representative results for the fidelity $F=\langle \Psi|\Phi\rangle$ between the AOD ansatz and the DMRG ground state $|\Phi\rangle$. In the delocalised phase we find that the DMRG results converge very rapidly, and that only the first few sites of the chain representation of the bath are appreciably excited. The fidelity with the AOD state (equivalent to the SH state) is extremely high considering the many-body nature of the system, and we conclude that the AOD/SH ansatz works well at weak coupling. As $\alpha$ approaches $\alpha_{c}$, the fidelity decreases considerably and is in fact zero above the transition. We believe that this suppression of $F$ may arise from the truncation of bosonic fock space used in the DMRG simulation and the finite size of the harmonic chain. 

Applying the unitary transformation which maps the SBM to a harmonic chain to the  AOD ansatz, one finds that that the average occupation of site $n$ in the chain $N_{av}(n)$ decreases very rapidly in the delocalised phase, and an accurate simulation of the ground state with a small $N_{b}$ and chain length is possible. However as $\alpha$ approaches $\alpha_{c}$, the decay of $N_{av}(n)$ along the chain becomes much weaker and contributions from many sites need to be included. Although $N_{av}(n)$ is bounded along the chain, the convergence w.r.t chain length in the DMRG simulations becomes much slower just below the transition, and for a fixed number of sites we observe a continuous cross-over rather than a sharp transition as $\alpha$ is increased. The magnetisation typically begins to appear in the DMRG simulation at $\alpha$ slightly below the $\alpha_{c}$ predicted from the AOD ansatz, with the deviation decreasing as the number of chain sites is increased. This leads to the reduction of the fidelity seen in Fig. (\ref{dmrg}) as the transition is approached from below. Above the $\alpha_{c}$ predicted from the AOD, the fidelity between DMRG and AOD groundstates becomes extremely small. This arises because $N_{av}(n)$ actually diverges along the chain in the magnetic phase. This can be shown directly using the chain mapping \cite{chinchain10}, which predicts that $N_{av}(n)\propto n^{1-2s}$ as $n\rightarrow\infty$ when $M$ is finite. As in NRG, the DMRG approach uses a finite number of bosonic fock states $N_{b}$ to represent each oscillator of the chain, and the divergence of the chain populations in the AOD ansatz for $s<0.5$ cannot be described in this truncated basis. Consequently, the overlap $\langle \Psi|\phi\rangle$ goes to zero exponentially with chain length, as the norm of the projection of the AOD ansatz onto the truncated fock space vanishes with increasing chain size. This rapid suppression of $F$ with chain size is shown in the inset of Fig. (\ref{dmrg}).  Although of a different nature, the use of a finite $N_{b}$ in the NRG description of the transition in $0<s<0.5$ also appears to cause problems, as discussed in Refs. \cite{hou10, vojta10}.
 
The result that $N_{av}(n)\propto n^{1-2s}$ suggests that the current application of DMRG is unable to decribe the localised phase for $s<1/2$. However, the AOD ansatz may provide a means to resolve this issue. The divergence of $N_{av}(n)$ is ultimately related to the infra-red divergence, which is itself a consequence of the appearance of the 'mean field' displacement of the oscillators above the transition. The diverging occupations arising from this displacement could be removed by a simple unitary transformation $H\rightarrow UHU^{-1}$, where $U=\exp\left(\frac{1}{2}M\sum_{l}g_{l}\omega_{l}^{-1}(a_{l}-a_{l}^{\dagger})\right)$. This corresponds to using a displaced oscillator basis where the effects of the mean-field displacement (computed from the AOD ansatz) are removed from the problem. Performing this unitary tranforamtion and then mapping the resulting SBM onto a chain, we find that $N_{av}(n)$ computed from the AOD ansatz now decreases along the chain at a similar rate to that found in the delocalised phase. Therefore, using the AOD prediction for $M$ and a unitary transormation may allow an efficient simulation of the localised phase using DMRG in a truncated bosonic fock space. In the likely event that the AOD magnetisation is not exactly the true ground state magnetisation of the model, the unitary transformation above will not completely cancel out the infra-red divergence and the siumlation will converge slowly.  Therefore, by varying the value of $M$ used in the unitary transformation, we can seek out the true value by looking for the value with the fastest convergence time. This idea of removing the effect of the mean-field displacment was orginally used in the SPSA study where it was found to yield an efficient and accurate simulation of the magnetic phase \cite{alvermann09}. However, this method will only be applicable to the case described here ($s<0.5$), where a mean-field bias arises from the bath. For $s>0.5$ the transition is not expected to be mean-field, and it is perhaps interesting that in our chain mapping the asymptotic values of $N_{av}(n)$  are qualitatively different above and below $s=0.5$. 

\section{Conclusion}
We have presented a strongly-correlated variational ansatz for the ground state of the sub-Ohmic spin-boson model and have found that it undergoes a continuous quantum phase transition at a critical coupling strength to the bath. The critical properties and the values of observables that it predicts are very good approximations to those found by numerical methods, and by an analysis of the microscopic structure of the ansatz we have been able to describe the rich physics of the adiabatic and non-adiabatic modes that may drive this transition. We have also found that an infra-red divegence of the low frequency bosons always accompanies the transition in this theory, and this may be of relevance for numerical approaches to this problem, as illustrated with our numerical DMRG simulation of the sub-Ohmic ground state.

\acknowledgements
We thank R. Bulla for discussions and the EU-STREPs CORNER and PICC as well as the Alexander von Humboldt foundation for support. J. P. was supported by the Fundacion Seneca Project No. 11920/PI/09-j, Ministerio  de Ciencia e Innovacion Project No. FIS2009-13483-C02-02 and HOPE-CSD2007-00007 (Consolider Ingenio).

\bibliography{sbvar}

\begin{thebibliography}{27}%
\makeatletter
\providecommand \@ifxundefined [1]{%
 \@ifx{#1\undefined}
}%
\providecommand \@ifnum [1]{%
 \ifnum #1\expandafter \@firstoftwo
 \else \expandafter \@secondoftwo
 \fi
}%
\providecommand \@ifx [1]{%
 \ifx #1\expandafter \@firstoftwo
 \else \expandafter \@secondoftwo
 \fi
}%
\providecommand \natexlab [1]{#1}%
\providecommand \enquote  [1]{``#1''}%
\providecommand \bibnamefont  [1]{#1}%
\providecommand \bibfnamefont [1]{#1}%
\providecommand \citenamefont [1]{#1}%
\providecommand \href@noop [0]{\@secondoftwo}%
\providecommand \href [0]{\begingroup \@sanitize@url \@href}%
\providecommand \@href[1]{\@@startlink{#1}\@@href}%
\providecommand \@@href[1]{\endgroup#1\@@endlink}%
\providecommand \@sanitize@url [0]{\catcode `\\12\catcode `\$12\catcode
  `\&12\catcode `\#12\catcode `\^12\catcode `\_12\catcode `\%12\relax}%
\providecommand \@@startlink[1]{}%
\providecommand \@@endlink[0]{}%
\providecommand \url  [0]{\begingroup\@sanitize@url \@url }%
\providecommand \@url [1]{\endgroup\@href {#1}{\urlprefix }}%
\providecommand \urlprefix  [0]{URL }%
\providecommand \Eprint [0]{\href }%
\providecommand \doibase [0]{http://dx.doi.org/}%
\providecommand \selectlanguage [0]{\@gobble}%
\providecommand \bibinfo  [0]{\@secondoftwo}%
\providecommand \bibfield  [0]{\@secondoftwo}%
\providecommand \translation [1]{[#1]}%
\providecommand \BibitemOpen [0]{}%
\providecommand \bibitemStop [0]{}%
\providecommand \bibitemNoStop [0]{.\EOS\space}%
\providecommand \EOS [0]{\spacefactor3000\relax}%
\providecommand \BibitemShut  [1]{\csname bibitem#1\endcsname}%
\let\auto@bib@innerbib\@empty
\bibitem [{\citenamefont {Faoro}\ and\ \citenamefont {Ioffe}(2007)}]{kit}%
  \BibitemOpen
  \bibfield  {author} {\bibinfo {author} {\bibfnamefont {L.}~\bibnamefont
  {Faoro}}\ and\ \bibinfo {author} {\bibfnamefont {L.~B.}\ \bibnamefont
  {Ioffe}},\ }\href {\doibase 10.1103/PhysRevB.75.132505} {\bibfield  {journal}
  {\bibinfo  {journal} {Phys. Rev. B}\ }\textbf {\bibinfo {volume} {75}},\
  \bibinfo {pages} {132505} (\bibinfo {year} {2007})}\BibitemShut {NoStop}%
\bibitem [{\citenamefont {Schriefl}\ \emph {et~al.}(2006)\citenamefont
  {Schriefl}, \citenamefont {Makhlin}, \citenamefont {Shnirman},\ and\
  \citenamefont {Schön}}]{kit2}%
  \BibitemOpen
  \bibfield  {author} {\bibinfo {author} {\bibfnamefont {J.}~\bibnamefont
  {Schriefl}}, \bibinfo {author} {\bibfnamefont {Y.}~\bibnamefont {Makhlin}},
  \bibinfo {author} {\bibfnamefont {A.}~\bibnamefont {Shnirman}}, \ and\
  \bibinfo {author} {\bibfnamefont {G.}~\bibnamefont {Schön}},\ }\href@noop {}
  {\bibfield  {journal} {\bibinfo  {journal} {New Journal of Physics}\ }\textbf
  {\bibinfo {volume} {8}},\ \bibinfo {pages} {1} (\bibinfo {year}
  {2006})}\BibitemShut {NoStop}%
\bibitem [{\citenamefont {Bulla}\ \emph {et~al.}(2008)\citenamefont {Bulla},
  \citenamefont {Costi},\ and\ \citenamefont {Pruschke}}]{bullarev08}%
  \BibitemOpen
  \bibfield  {author} {\bibinfo {author} {\bibfnamefont {R.}~\bibnamefont
  {Bulla}}, \bibinfo {author} {\bibfnamefont {T.~A.}\ \bibnamefont {Costi}}, \
  and\ \bibinfo {author} {\bibfnamefont {T.}~\bibnamefont {Pruschke}},\ }\href
  {\doibase 10.1103/RevModPhys.80.395} {\bibfield  {journal} {\bibinfo
  {journal} {Rev. Mod. Phys.}\ }\textbf {\bibinfo {volume} {80}},\ \bibinfo
  {pages} {395} (\bibinfo {year} {2008})}\BibitemShut {NoStop}%
\bibitem [{\citenamefont {Mohseni}\ \emph {et~al.}(2008)\citenamefont
  {Mohseni}, \citenamefont {Rebentrost}, \citenamefont {Lloyd},\ and\
  \citenamefont {Aspuru-Guzik}}]{nat2}%
  \BibitemOpen
  \bibfield  {author} {\bibinfo {author} {\bibfnamefont {M.}~\bibnamefont
  {Mohseni}}, \bibinfo {author} {\bibfnamefont {P.}~\bibnamefont {Rebentrost}},
  \bibinfo {author} {\bibfnamefont {S.}~\bibnamefont {Lloyd}}, \ and\ \bibinfo
  {author} {\bibfnamefont {A.}~\bibnamefont {Aspuru-Guzik}},\ }\href {\doibase
  10.1063/1.3002335} {\bibfield  {journal} {\bibinfo  {journal} {The Journal of
  Chemical Physics}\ }\textbf {\bibinfo {volume} {129}},\ \bibinfo {eid}
  {174106} (\bibinfo {year} {2008})}\BibitemShut {NoStop}%
\bibitem [{\citenamefont {Plenio}\ and\ \citenamefont
  {Huelga}(2008)}]{plenio08}%
  \BibitemOpen
  \bibfield  {author} {\bibinfo {author} {\bibfnamefont {M.~B.}\ \bibnamefont
  {Plenio}}\ and\ \bibinfo {author} {\bibfnamefont {S.~F.}\ \bibnamefont
  {Huelga}},\ }\href@noop {} {\bibfield  {journal} {\bibinfo  {journal} {New J.
  Phys.}\ }\textbf {\bibinfo {volume} {10}},\ \bibinfo {pages} {113019}
  (\bibinfo {year} {2008})}\BibitemShut {NoStop}%
\bibitem [{\citenamefont {Chin}\ \emph
  {et~al.}(2010{\natexlab{a}})\citenamefont {Chin}, \citenamefont {Datta},
  \citenamefont {Caruso}, \citenamefont {Huelga},\ and\ \citenamefont
  {Plenio}}]{chinnat10}%
  \BibitemOpen
  \bibfield  {author} {\bibinfo {author} {\bibfnamefont {A.~W.}\ \bibnamefont
  {Chin}}, \bibinfo {author} {\bibfnamefont {A.}~\bibnamefont {Datta}},
  \bibinfo {author} {\bibfnamefont {F.}~\bibnamefont {Caruso}}, \bibinfo
  {author} {\bibfnamefont {S.~F.}\ \bibnamefont {Huelga}}, \ and\ \bibinfo
  {author} {\bibfnamefont {M.~B.}\ \bibnamefont {Plenio}},\ }\href@noop {}
  {\bibfield  {journal} {\bibinfo  {journal} {New Journal of Physics}\ }\textbf
  {\bibinfo {volume} {12}},\ \bibinfo {pages} {065002} (\bibinfo {year}
  {2010}{\natexlab{a}})}\BibitemShut {NoStop}%
\bibitem [{\citenamefont {Caruso}\ \emph {et~al.}(2009)\citenamefont {Caruso},
  \citenamefont {Alex W.~Chin}, \citenamefont {Datta}, \citenamefont {Huelga},\
  and\ \citenamefont {Plenio}}]{caruso09}%
  \BibitemOpen
  \bibfield  {author} {\bibinfo {author} {\bibfnamefont {F.}~\bibnamefont
  {Caruso}}, \bibinfo {author} {\bibfnamefont {A.~W.}\ \bibnamefont {Alex
  W.~Chin}}, \bibinfo {author} {\bibfnamefont {A.}~\bibnamefont {Datta}},
  \bibinfo {author} {\bibfnamefont {S.~F.}\ \bibnamefont {Huelga}}, \ and\
  \bibinfo {author} {\bibfnamefont {M.~B.}\ \bibnamefont {Plenio}},\
  }\href@noop {} {\bibfield  {journal} {\bibinfo  {journal} {J. Chem. Phys}\
  }\textbf {\bibinfo {volume} {131}},\ \bibinfo {pages} {105106} (\bibinfo
  {year} {2009})}\BibitemShut {NoStop}%
\bibitem [{\citenamefont {Prior}\ \emph {et~al.}(2010)\citenamefont {Prior},
  \citenamefont {Chin}, \citenamefont {Huelga},\ and\ \citenamefont
  {Plenio}}]{prior10}%
  \BibitemOpen
  \bibfield  {author} {\bibinfo {author} {\bibfnamefont {J.}~\bibnamefont
  {Prior}}, \bibinfo {author} {\bibfnamefont {A.~W.}\ \bibnamefont {Chin}},
  \bibinfo {author} {\bibfnamefont {S.~F.}\ \bibnamefont {Huelga}}, \ and\
  \bibinfo {author} {\bibfnamefont {M.~B.}\ \bibnamefont {Plenio}},\ }\href
  {\doibase 10.1103/PhysRevLett.105.050404} {\bibfield  {journal} {\bibinfo
  {journal} {Phys. Rev. Lett.}\ }\textbf {\bibinfo {volume} {105}},\ \bibinfo
  {pages} {050404} (\bibinfo {year} {2010})}\BibitemShut {NoStop}%
\bibitem [{\citenamefont {Leggett}\ \emph {et~al.}(1987)\citenamefont
  {Leggett}, \citenamefont {Chakravarty}, \citenamefont {Dorsey}, \citenamefont
  {Fisher}, \citenamefont {Garg},\ and\ \citenamefont {Zwerger}}]{leggett}%
  \BibitemOpen
  \bibfield  {author} {\bibinfo {author} {\bibfnamefont {A.~J.}\ \bibnamefont
  {Leggett}}, \bibinfo {author} {\bibfnamefont {S.}~\bibnamefont
  {Chakravarty}}, \bibinfo {author} {\bibfnamefont {A.~T.}\ \bibnamefont
  {Dorsey}}, \bibinfo {author} {\bibfnamefont {M.~P.~A.}\ \bibnamefont
  {Fisher}}, \bibinfo {author} {\bibfnamefont {A.}~\bibnamefont {Garg}}, \ and\
  \bibinfo {author} {\bibfnamefont {W.}~\bibnamefont {Zwerger}},\ }\href
  {\doibase 10.1103/RevModPhys.59.1} {\bibfield  {journal} {\bibinfo  {journal}
  {Rev. Mod. Phys.}\ }\textbf {\bibinfo {volume} {59}},\ \bibinfo {pages} {1}
  (\bibinfo {year} {1987})}\BibitemShut {NoStop}%
\bibitem [{\citenamefont {Weiss}(1993)}]{weiss}%
  \BibitemOpen
  \bibfield  {author} {\bibinfo {author} {\bibfnamefont {U.}~\bibnamefont
  {Weiss}},\ }\href@noop {} {\emph {\bibinfo {title} {Quantum Dissipative
  Systems}}}\ (\bibinfo  {publisher} {World Scientific, Singapore},\ \bibinfo
  {year} {1993})\BibitemShut {NoStop}%
\bibitem [{\citenamefont {Vojta}\ \emph {et~al.}(2005)\citenamefont {Vojta},
  \citenamefont {Tong},\ and\ \citenamefont {Bulla}}]{vojta05}%
  \BibitemOpen
  \bibfield  {author} {\bibinfo {author} {\bibfnamefont {M.}~\bibnamefont
  {Vojta}}, \bibinfo {author} {\bibfnamefont {N.-H.}\ \bibnamefont {Tong}}, \
  and\ \bibinfo {author} {\bibfnamefont {R.}~\bibnamefont {Bulla}},\ }\href
  {\doibase 10.1103/PhysRevLett.94.070604} {\bibfield  {journal} {\bibinfo
  {journal} {Phys. Rev. Lett.}\ }\textbf {\bibinfo {volume} {94}},\ \bibinfo
  {pages} {070604} (\bibinfo {year} {2005})}\BibitemShut {NoStop}%
\bibitem [{\citenamefont {Kirchner}\ \emph {et~al.}(2009)\citenamefont
  {Kirchner}, \citenamefont {Si},\ and\ \citenamefont
  {Ingersent}}]{kirchner09}%
  \BibitemOpen
  \bibfield  {author} {\bibinfo {author} {\bibfnamefont {S.}~\bibnamefont
  {Kirchner}}, \bibinfo {author} {\bibfnamefont {Q.}~\bibnamefont {Si}}, \ and\
  \bibinfo {author} {\bibfnamefont {K.}~\bibnamefont {Ingersent}},\ }\href
  {\doibase 10.1103/PhysRevLett.102.166405} {\bibfield  {journal} {\bibinfo
  {journal} {Phys. Rev. Lett.}\ }\textbf {\bibinfo {volume} {102}},\ \bibinfo
  {pages} {166405} (\bibinfo {year} {2009})}\BibitemShut {NoStop}%
\bibitem [{\citenamefont {Kirchner}(2010)}]{kirchner10}%
  \BibitemOpen
  \bibfield  {author} {\bibinfo {author} {\bibfnamefont {S.}~\bibnamefont
  {Kirchner}},\ }\href@noop {} {\bibfield  {journal} {\bibinfo  {journal}
  {arxiv:1007.4558}\ } (\bibinfo {year} {2010})}\BibitemShut {NoStop}%
\bibitem [{\citenamefont {Vojta}\ \emph {et~al.}(2010)\citenamefont {Vojta},
  \citenamefont {Bulla}, \citenamefont {G\"uttge},\ and\ \citenamefont
  {Anders}}]{vojta10}%
  \BibitemOpen
  \bibfield  {author} {\bibinfo {author} {\bibfnamefont {M.}~\bibnamefont
  {Vojta}}, \bibinfo {author} {\bibfnamefont {R.}~\bibnamefont {Bulla}},
  \bibinfo {author} {\bibfnamefont {F.}~\bibnamefont {G\"uttge}}, \ and\
  \bibinfo {author} {\bibfnamefont {F.}~\bibnamefont {Anders}},\ }\href
  {\doibase 10.1103/PhysRevB.81.075122} {\bibfield  {journal} {\bibinfo
  {journal} {Phys. Rev. B}\ }\textbf {\bibinfo {volume} {81}},\ \bibinfo
  {pages} {075122} (\bibinfo {year} {2010})}\BibitemShut {NoStop}%
\bibitem [{\citenamefont {Hou}\ and\ \citenamefont {Tong}(2010)}]{hou10}%
  \BibitemOpen
  \bibfield  {author} {\bibinfo {author} {\bibfnamefont {Z.~H.}\ \bibnamefont
  {Hou}}\ and\ \bibinfo {author} {\bibfnamefont {N.~H.}\ \bibnamefont {Tong}},\
  }\href@noop {} {\bibfield  {journal} {\bibinfo  {journal} {Eur. Phys. J. B.}\
  } (\bibinfo {year} {2010})}\BibitemShut {NoStop}%
\bibitem [{\citenamefont {Winter}\ \emph {et~al.}(2009)\citenamefont {Winter},
  \citenamefont {Rieger}, \citenamefont {Vojta},\ and\ \citenamefont
  {Bulla}}]{winter09}%
  \BibitemOpen
  \bibfield  {author} {\bibinfo {author} {\bibfnamefont {A.}~\bibnamefont
  {Winter}}, \bibinfo {author} {\bibfnamefont {H.}~\bibnamefont {Rieger}},
  \bibinfo {author} {\bibfnamefont {M.}~\bibnamefont {Vojta}}, \ and\ \bibinfo
  {author} {\bibfnamefont {R.}~\bibnamefont {Bulla}},\ }\href {\doibase
  10.1103/PhysRevLett.102.030601} {\bibfield  {journal} {\bibinfo  {journal}
  {Phys. Rev. Lett.}\ }\textbf {\bibinfo {volume} {102}},\ \bibinfo {pages}
  {030601} (\bibinfo {year} {2009})}\BibitemShut {NoStop}%
\bibitem [{\citenamefont {Alvermann}\ and\ \citenamefont
  {Fehske}(2009)}]{alvermann09}%
  \BibitemOpen
  \bibfield  {author} {\bibinfo {author} {\bibfnamefont {A.}~\bibnamefont
  {Alvermann}}\ and\ \bibinfo {author} {\bibfnamefont {H.}~\bibnamefont
  {Fehske}},\ }\href {\doibase 10.1103/PhysRevLett.102.150601} {\bibfield
  {journal} {\bibinfo  {journal} {Phys. Rev. Lett.}\ }\textbf {\bibinfo
  {volume} {102}},\ \bibinfo {pages} {150601} (\bibinfo {year}
  {2009})}\BibitemShut {NoStop}%
\bibitem [{\citenamefont {Zhang}\ \emph {et~al.}(2010)\citenamefont {Zhang},
  \citenamefont {Chen},\ and\ \citenamefont {Wang}}]{zhang10}%
  \BibitemOpen
  \bibfield  {author} {\bibinfo {author} {\bibfnamefont {Y.-Y.}\ \bibnamefont
  {Zhang}}, \bibinfo {author} {\bibfnamefont {Q.-H.}\ \bibnamefont {Chen}}, \
  and\ \bibinfo {author} {\bibfnamefont {K.-L.}\ \bibnamefont {Wang}},\ }\href
  {\doibase 10.1103/PhysRevB.81.121105} {\bibfield  {journal} {\bibinfo
  {journal} {Phys. Rev. B}\ }\textbf {\bibinfo {volume} {81}},\ \bibinfo
  {pages} {121105} (\bibinfo {year} {2010})}\BibitemShut {NoStop}%
\bibitem [{\citenamefont {Chin}\ \emph
  {et~al.}(2010{\natexlab{b}})\citenamefont {Chin}, \citenamefont {Rivas},
  \citenamefont {Huelga},\ and\ \citenamefont {Plenio}}]{chinchain10}%
  \BibitemOpen
  \bibfield  {author} {\bibinfo {author} {\bibfnamefont {A.~W.}\ \bibnamefont
  {Chin}}, \bibinfo {author} {\bibfnamefont {A.}~\bibnamefont {Rivas}},
  \bibinfo {author} {\bibfnamefont {S.~F.}\ \bibnamefont {Huelga}}, \ and\
  \bibinfo {author} {\bibfnamefont {M.~B.}\ \bibnamefont {Plenio}},\
  }\href@noop {} {\bibfield  {journal} {\bibinfo  {journal} {J. Math. Phys}\
  }\textbf {\bibinfo {volume} {51}},\ \bibinfo {pages} {092109} (\bibinfo
  {year} {2010}{\natexlab{b}})}\BibitemShut {NoStop}%
\bibitem [{\citenamefont {Bulla}\ \emph {et~al.}(2005)\citenamefont {Bulla},
  \citenamefont {Lee}, \citenamefont {Tong},\ and\ \citenamefont
  {Vojta}}]{Bulla05}%
  \BibitemOpen
  \bibfield  {author} {\bibinfo {author} {\bibfnamefont {R.}~\bibnamefont
  {Bulla}}, \bibinfo {author} {\bibfnamefont {H.-J.}\ \bibnamefont {Lee}},
  \bibinfo {author} {\bibfnamefont {N.-H.}\ \bibnamefont {Tong}}, \ and\
  \bibinfo {author} {\bibfnamefont {M.}~\bibnamefont {Vojta}},\ }\href
  {\doibase 10.1103/PhysRevB.71.045122} {\bibfield  {journal} {\bibinfo
  {journal} {Phys. Rev. B}\ }\textbf {\bibinfo {volume} {71}},\ \bibinfo
  {pages} {045122} (\bibinfo {year} {2005})}\BibitemShut {NoStop}%
\bibitem [{\citenamefont {Silbey}\ and\ \citenamefont {Harris}(1984)}]{silbey}%
  \BibitemOpen
  \bibfield  {author} {\bibinfo {author} {\bibfnamefont {R.}~\bibnamefont
  {Silbey}}\ and\ \bibinfo {author} {\bibfnamefont {R.}~\bibnamefont
  {Harris}},\ }\href@noop {} {\bibfield  {journal} {\bibinfo  {journal} {J.
  Chem. Phys}\ }\textbf {\bibinfo {volume} {80}},\ \bibinfo {pages} {2615}
  (\bibinfo {year} {1984})}\BibitemShut {NoStop}%
\bibitem [{\citenamefont {Chin}\ and\ \citenamefont {Turlakov}(2006)}]{chin06}%
  \BibitemOpen
  \bibfield  {author} {\bibinfo {author} {\bibfnamefont {A.}~\bibnamefont
  {Chin}}\ and\ \bibinfo {author} {\bibfnamefont {M.}~\bibnamefont
  {Turlakov}},\ }\href {\doibase 10.1103/PhysRevB.73.075311} {\bibfield
  {journal} {\bibinfo  {journal} {Phys. Rev. B}\ }\textbf {\bibinfo {volume}
  {73}},\ \bibinfo {pages} {075311} (\bibinfo {year} {2006})}\BibitemShut
  {NoStop}%
\bibitem [{\citenamefont {L\"{u}}\ and\ \citenamefont {Zheng}(2007)}]{lu07}%
  \BibitemOpen
  \bibfield  {author} {\bibinfo {author} {\bibfnamefont {Z.}~\bibnamefont
  {L\"{u}}}\ and\ \bibinfo {author} {\bibfnamefont {H.}~\bibnamefont {Zheng}},\
  }\href {\doibase 10.1103/PhysRevB.75.054302} {\bibfield  {journal} {\bibinfo
  {journal} {Phys. Rev. B}\ }\textbf {\bibinfo {volume} {75}},\ \bibinfo {eid}
  {054302} (\bibinfo {year} {2007})}\BibitemShut {NoStop}%
\bibitem [{\citenamefont {Corless}\ \emph {et~al.}(1996)\citenamefont
  {Corless}, \citenamefont {Gonnet}, \citenamefont {Hare}, \citenamefont
  {Jeffrey},\ and\ \citenamefont {Knuth}}]{lambert}%
  \BibitemOpen
  \bibfield  {author} {\bibinfo {author} {\bibfnamefont {R.~M.}\ \bibnamefont
  {Corless}}, \bibinfo {author} {\bibfnamefont {G.~H.}\ \bibnamefont {Gonnet}},
  \bibinfo {author} {\bibfnamefont {D.}~\bibnamefont {Hare}}, \bibinfo {author}
  {\bibfnamefont {D.~J.}\ \bibnamefont {Jeffrey}}, \ and\ \bibinfo {author}
  {\bibfnamefont {D.}~\bibnamefont {Knuth}},\ }\href@noop {} {\bibfield
  {journal} {\bibinfo  {journal} {Adv. Comput. Math}\ }\textbf {\bibinfo
  {volume} {5}},\ \bibinfo {pages} {326} (\bibinfo {year} {1996})}\BibitemShut
  {NoStop}%
\bibitem [{\citenamefont {Anders}\ \emph {et~al.}(2007)\citenamefont {Anders},
  \citenamefont {Bulla},\ and\ \citenamefont {Vojta}}]{anders07}%
  \BibitemOpen
  \bibfield  {author} {\bibinfo {author} {\bibfnamefont {F.~B.}\ \bibnamefont
  {Anders}}, \bibinfo {author} {\bibfnamefont {R.}~\bibnamefont {Bulla}}, \
  and\ \bibinfo {author} {\bibfnamefont {M.}~\bibnamefont {Vojta}},\ }\href
  {\doibase 10.1103/PhysRevLett.98.210402} {\bibfield  {journal} {\bibinfo
  {journal} {Phys. Rev. Lett.}\ }\textbf {\bibinfo {volume} {98}},\ \bibinfo
  {pages} {210402} (\bibinfo {year} {2007})}\BibitemShut {NoStop}%
\bibitem [{\citenamefont {Le~Hur}\ \emph {et~al.}(2007)\citenamefont {Le~Hur},
  \citenamefont {Doucet-Beaupr\'e},\ and\ \citenamefont
  {Hofstetter}}]{lehur07}%
  \BibitemOpen
  \bibfield  {author} {\bibinfo {author} {\bibfnamefont {K.}~\bibnamefont
  {Le~Hur}}, \bibinfo {author} {\bibfnamefont {P.}~\bibnamefont
  {Doucet-Beaupr\'e}}, \ and\ \bibinfo {author} {\bibfnamefont
  {W.}~\bibnamefont {Hofstetter}},\ }\href {\doibase
  10.1103/PhysRevLett.99.126801} {\bibfield  {journal} {\bibinfo  {journal}
  {Phys. Rev. Lett.}\ }\textbf {\bibinfo {volume} {99}},\ \bibinfo {pages}
  {126801} (\bibinfo {year} {2007})}\BibitemShut {NoStop}%
\bibitem [{\citenamefont {Kehrein}\ and\ \citenamefont
  {Mielke}(1996)}]{kehrein}%
  \BibitemOpen
  \bibfield  {author} {\bibinfo {author} {\bibfnamefont {S.~K.}\ \bibnamefont
  {Kehrein}}\ and\ \bibinfo {author} {\bibfnamefont {A.}~\bibnamefont
  {Mielke}},\ }\href@noop {} {\bibfield  {journal} {\bibinfo  {journal} {Phys.
  Lett. A}\ }\textbf {\bibinfo {volume} {219}},\ \bibinfo {pages} {313}
  (\bibinfo {year} {1996})}\BibitemShut {NoStop}%
\end{thebibliography}%

\end{document}